\documentclass[twocolumn,showpacs,preprintnumbers, reprint, amsmath,
amssymb,prl]{revtex4-1}

\usepackage{graphicx}
\usepackage{dcolumn}
\usepackage{bm}

\graphicspath{{./figs/}}

\newcommand{\bodyvel}{{\boldsymbol{\xi}}}				
\newcommand{\beq}{\begin{equation}}
\newcommand{\eeq}{\end{equation}}
\newcommand{\mixedconn}{\mathbf{A}}		

\newcommand{\subletter}[1]{\emph{#1}}

\begin{document}
\title{Geometric visualization of self-propulsion in a complex medium}
\author{Ross L. Hatton}
\affiliation{School of Mechanical, Industrial, and Manufacturing Engineering, Oregon State University, Corvallis, OR 97331
}

\author{Howie Choset}
\affiliation{Robotics Institute, Carnegie Mellon University, Pittsburgh, PA 15213
}

\author{Yang Ding}
\affiliation{School of Physics, Georgia Institute of Technology,
Atlanta, Georgia 30332}

\author{Daniel I.\ Goldman*}
\affiliation{School of Physics, Georgia Institute of Technology,
Atlanta, Georgia 30332}
\date{\today}

\begin{abstract} 

Combining geometric mechanics theory, laboratory robotic experiment and numerical simulation, we study the locomotion in granular media (GM) of the simplest non-inertial swimmer, the Purcell three-link swimmer. Using granular resistive force laws as inputs, the theory relates translation and rotation of the body to shape changes (movements of the links). This allows analysis, visualization, and prediction of effective movements that are verified by experiment. The geometric approach also facilitates comparison between swimming in GM and in viscous fluids.

\end{abstract}

\maketitle
{\em Introduction}--Locomotion of animals and robots emerges through the interplay of body deformations coupled to an environment. Finding this relationship is often a challenge: for example, in Newtonian fluids, although researchers have long analyzed~\cite{childressbook,vogelbook,lauApow2009} the Navier-Stokes equations and simpler representations~\cite{wangreview} to gain insight into flight and swimming, analytic investigation is often impossible, and high fidelity approximations are computationally costly. Studying the motion of organisms~\cite{mosauer1932adaptive} and robots~\cite{synthsandswim} that maneuver through complex environments like sand, rubble, and debris, and microscopic organisms~\cite{lauApow2009} that move through complex biomaterials can be even more complicated---often such materials are not even described by equations at the level of Navier-Stokes.

Certain kinds of movement are kinematic, in that the net displacement is a function of the deformation and is independent of its rate. Taking advantage of this property for low Reynolds number (Re) swimming in viscous Newtonian fluids, Shapere and Wilczek~\cite{Shapere:1989} introduced a geometric approach using the notion of {gauge symmetries}, which are equivalencies in the system dynamics across different configurations. These symmetries reduce the effective dimensionality of the system and facilitate interpretation of  dynamics in terms of geometric concepts such as areas, lengths, and curvatures. The geometric approach has been further developed~\cite{ostrowski98a,Melli:2006} and now enables evaluations~\cite{Hatton:2011IJRR} of systems' locomotion capabilities in the form of low-dimensional, readily visualizable representations of the system's motion for any gait (cyclic change in body shape). These tools allow useful gaits to be identified by inspection, without costly trial-and-error optimization~\cite{tamAhos}. However, the insights afforded by these geometric tools have been restricted to systems---including  viscous swimmers~\cite{Shapere:1989, Melli:2006,Hatton:2010ASME, Hatton:2011RSS} and planar reorienting satellites~\cite{walsh95}---with analytically-describable linear dynamics.

\begin{figure}[h]
\includegraphics[width={1\hsize}]{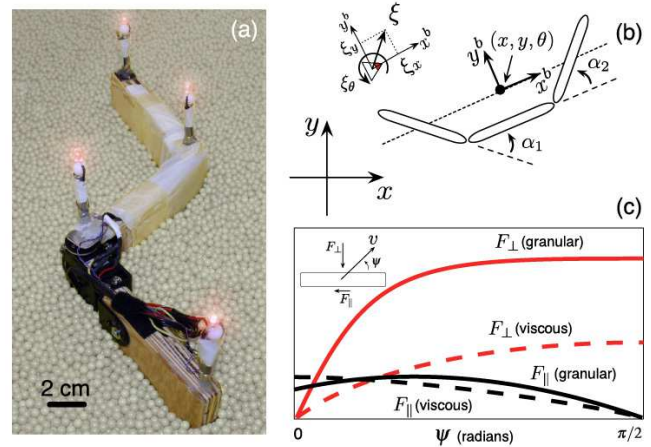}
\caption{Three-link swimmers: (A) The robot resting on a granular medium (GM) of plastic spheres. The lit masts at the ends of the links are markers for tracking the robot when its body is submerged. The rubber skin on the front section has been removed to display the mechanical structure. Each link is 5.4 $\times$ 2.8 $\times$ 14.7 cm$^3$. (B) Analytical three-link model with a body frame corresponding to a weighted average of the link positions and orientations. (c) Force relations for a rod dragged in GM (adapted from ~\cite{maladen2009undulatory}) [solid] and a long slender rod dragged in a low Re fluid [dashed]. Scaling is chosen such that the maximum values of $F_{\parallel}$ are equal for the GM and low Re systems.}
\label{fig:sandfishthreelink}
\end{figure}
Previously, we studied~\cite{maladen2009undulatory} the sand-swimming of a sandfish lizard and a robot model in dry granular media (GM), arguably the simplest flowing terrestrial material. In theory, simulation and experiment, we demonstrated that the GM surrounding an immersed undulatory swimmer could be modeled as a ``frictional fluid" in which forces are dominated by Coulomb friction, making them insensitive to rate and in which inertial effects are small. However, our analysis could only hint at the range of behaviors possible in sand-swimming. Here we demonstrate the efficacy of the geometric approach to reveal principles of swimming in GM, despite a lack of fundamental equations of motion. We empirically generate a geometric swimming model in GM for the three-link swimmer (Fig.~\ref{fig:sandfishthreelink}) first introduced by Purcell~\cite{Purcell:1977} as a simple swimmer to study locomotion in viscous fluids~\cite{becAkoe2003}. We use this model to analyze different locomotor behaviors and to compare swimming in GM to swimming in a viscous fluid.

{\em Geometric mechanics, the resistive force model, and the three link swimmer}--The key ingredient in applying geometric theory to motion in GM is the ansatz that at any given shape  (with joint angles specified by the vector $\boldsymbol{\alpha}=(\alpha_1,\alpha_2)$), the swimmer's body velocity $\bodyvel$ is linearly proportional to its shape velocity ($\dot{\boldsymbol{\alpha}}$), such that the relationship between shape, shape velocity, and body velocity can be expressed as
\begin{equation}
\bodyvel = \mixedconn(\boldsymbol{\alpha}) \cdot \dot{\boldsymbol{\alpha}} = (\mixedconn^x(\boldsymbol{\alpha}) \cdot \dot{\boldsymbol{\alpha}},\mixedconn^y(\boldsymbol{\alpha}) \cdot \dot{\boldsymbol{\alpha}},\mixedconn^\theta(\boldsymbol{\alpha}) \cdot \dot{\boldsymbol{\alpha}}),
\label{eq:kinrecon}
\end{equation}
where $\mixedconn({\boldsymbol{\alpha}})$ is referred to as the \emph{local connection} (or Jacobian) matrix~\cite{Bloch:03}.  Local connection models have been identified for diverse locomotion modes~\cite{Shapere:1989,walsh95,ostrowski98a, Bloch:03,Melli:2006,Shammas:2007,Avron:2008,Hatton:2010ASME,Kelly:SIAM}, in particular for swimmers in low Re ~\cite{Shapere:1989,Avron:2008,Hatton:2010ASME,Kelly:SIAM}, viscous environments that qualitatively resemble those seen in granular swimming~\cite{maladen2009undulatory}. The existence of such a model for motion in GM is further suggested by our previous results~\cite{maladen2009undulatory} showing that sand-swimming is kinematic. Local linearity between shape and position velocities sufficies to produce kinematic motion, and is only slightly stronger than the necessary condition, proportionality between body and shape velocities.

GM lack equations equivalent to Navier-Stokes, so analytic derivations of the local connection used previously are not applicable. In their place, we have developed a numerical means of identifying $\mixedconn$, based on our empirically obtained granular resistive force laws~\cite{maladen2009undulatory,synthsandswim} and the observation that inertial forces on a low-speed swimmer are sufficiently small that the swimmer moves quasi-statically. The force laws (Fig.~\ref{fig:sandfishthreelink}) resemble those in low Re fluids~\cite{Taylor:1951}, although forces perpendicular to body segments are enhanced relative to those in true fluids. Integrating these forces along the swimmer's body at different $\mathbf{\alpha}$, $\mathbf{\dot{\alpha}}$, and $\bodyvel$ combinations and solving for the combinations that give force equilibria yields mappings from shape velocity to body velocity at each shape.



\begin{figure}
\includegraphics[width={1\hsize}]{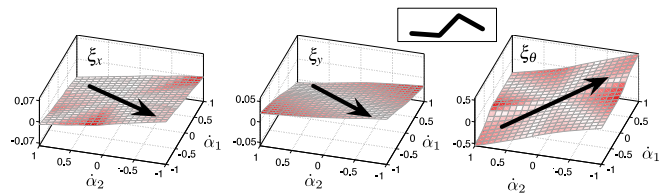}
\caption{The components of $\xi$ for a particular initial shape (inset), with $\alpha=(0.85,-1.14)$ radians; velocities are given in terms of body-lengths ($\xi_x$ and $\xi_y$) or radians ($\xi_theta$) per second. The components of the equilibrium $\xi$ (shown here for a body frame attached to the middle link) are almost linear functions of the shape velocity, and so can be characterized by the gradient vectors of the best-fit planes (black arrows). The color map shows the agreement between the equilibrium $\bodyvel$ and the fitted plane, with light regions exhibiting the least error.}
\label{fig:planes}
\end{figure}


\begin{figure}[h]
\includegraphics[width={1\hsize}]{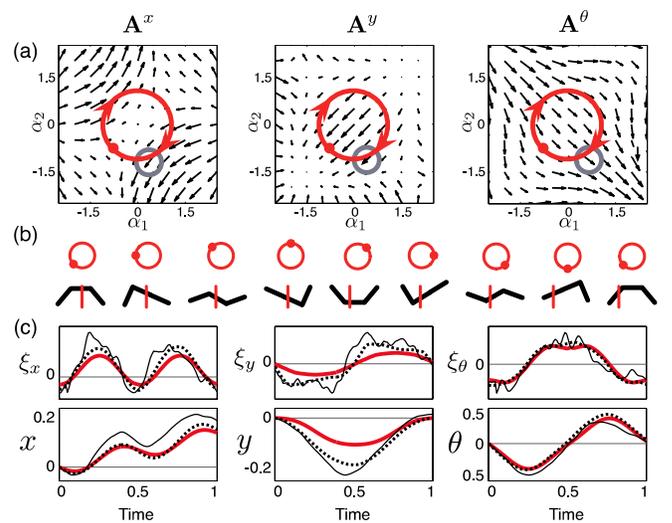}
\caption{(A) The local connection vector fields (Eqn.~1) $\mixedconn^x, \mixedconn^y,\mixedconn^\theta$. The red path through the shape space denotes a circle gait, and the grey circle identifies the gradients taken from Fig. 2. (B) Sequence of shapes and displacements along the gait shown above. The vertical red line is a reference position for displacement. (C) Along the gait, the $\bodyvel$ and net displacement predicted by the linear model (thick red line), the DEM simulation (dotted line) and robot experiments (thin black line) for a representative experiment at frequency 0.5 Hz. Lengths and angles are given in terms of body-lengths and radians, respectively. Time is in fractions of a gait cycle.}
\label{fig:vecheight}
\end{figure}

We find that across the space of shapes, graphs of the components of $\bodyvel$ as functions of $\dot{\alpha}$ are nearly planar, as in Fig.~\ref{fig:planes}. This planarity means that, despite the lack of an analytic linear relationship between shape and body velocities, there exists an implicit linear relationship to which we can fit a local connection of the form in Eqn.~1. Each row of $\mathbf{A(\alpha)}$ is then obtained by finding the best-fit planes for each component of the body velocity at each shape $\alpha$~\footnote{Across the $\pm 2.5$ radian domain of joint angles, the fits had $R^2$ values of $0.89\pm0.08$, $0.91\pm0.08$, and $0.91\pm0.14$ for the $x$, $y$, and $\theta$ rows of the local connection, respectively.}.  Taking these components over the set of shapes, we visualize each component of $\mixedconn$ as a vector field on the shape space (Fig. 3a), termed a \emph{connection vector field}.

{\em Testing the linear theory--} To test if the linear approximation (local connection approach) of the system dynamics can accurately compute movement over different gaits, we constructed a three-link robot in both experiment and multi-particle discrete element method (DEM) simulations, using identical techniques and parameters to those reported in~\cite{synthsandswim}. The robot (Fig. 1) consists of three wooden segments connected by two servomotors (Hitec, HSR 5980SG); the total mass of the robot is 0.56 kg. The segments are covered by a thin latex sleeve giving the robot a body-particle coefficient of friction of $0.4$. The robot is fully immersed in a large bed of $5.87\pm0.06$~mm diameter plastic spheres at a depth of $5.5$~cm from the top of the robot. Motion was tracked through a camera via the position of LED masts. Data was collected and averaged over 3 runs. The DEM simulation used $6$ mm diameter particles and a particle-particle and particle-body collision model incorporating Hertzian contact, normal dissipation and tangential Coulomb friction, with parameters previously validated against experiment~\cite{synthsandswim}.

Integrating the linear model's prediction of the swimmer's motion during a shape change is equivalent to taking line integrals on the connection vector fields along paths the system traces. For example, consider a gait in which the joints oscillate sinusoidally with a quarter-phase offset, producing a traveling wave of deformation along the body. This gait traces out a circle in the space of joint angles, shown in Fig.~\ref{fig:vecheight}\subletter{A}, generating positive and negative $\bodyvel$ as it flows along and against the vector fields, depicted in Fig.~\ref{fig:vecheight}\subletter{C}; these velocities can then be integrated into net displacements relative to the starting body frame. As illustrated in Fig.~\ref{fig:vecheight}\subletter{C}, the $x$ and $\theta$ components of the velocities and integrated positions predicted by the linear model agree with those found in experiment and DEM. The $y$ component of the velocity and displacement from the linear model qualitatively agree with those from the experiment and DEM but the magnitudes are reduced. We will return to this point.

\begin{figure}[h]
\includegraphics[width={1\hsize}]{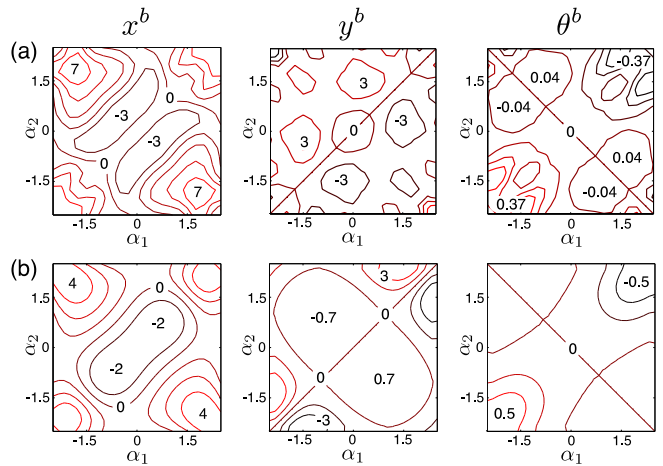}
\caption{Fundamental geometric diagrams, the constraint curvature functions (CCFs) for the (A) granular three-link swimmer and (B) low Re swimmer of equivalent dimensions (bottom). Each swimmer's body frame is optimized to a weighted-average of the three link frames, see~\cite{Hatton:2011IROS}. The units of the field values are $\text{body-lengths}/\text{joint-radians}^2$ for $x^b$ and $y^b$ and $\text{orientation-radians}/\text{joint-radians}^2$ for $\theta^b$. Values on the $x^b$ and $y^b$ fields have been multiplied by 100.}
\label{fig:geom}
\end{figure}

{\em Constraint curvature functions--}The local connection model of granular swimming simplifies evaluation of the displacement produced by a gait, but its greater usefulness derives from the ability to calculate the associated \emph{constraint curvature functions} (CCFs). These functions (calculated following the procedure in~\cite{Melli:2006,Avron:2008}) are plotted for GM in Fig.~\ref{fig:geom}\subletter{A}.  CCFs are closely related to the curls of the connection vector fields. In an extension of Stokes theorem, this curl-like nature means that the net displacement induced by a gait corresponds to the area integrals of the CCFs over the region the gait encloses in the shape space. Stokes theorem only applies for systems where the integrations are commutative and so does not directly apply to our swimmers, for which body-frame translations and rotations do not commute. CCFs circumvent this limitation by augmenting the curl with a Lie bracket term that linearly approximates the effects of noncommutativity. We introduced~\cite{Hatton:2011IJRR} coordinate optimization techniques that minimize the error in this linearization; for the granular-swimmer considered here (as in~\cite{Hatton:2011IJRR} and~\cite{Hatton:2013TRO:Swimming}) the error is negligible.



\begin{figure}[h]
\includegraphics[width={1\hsize}]{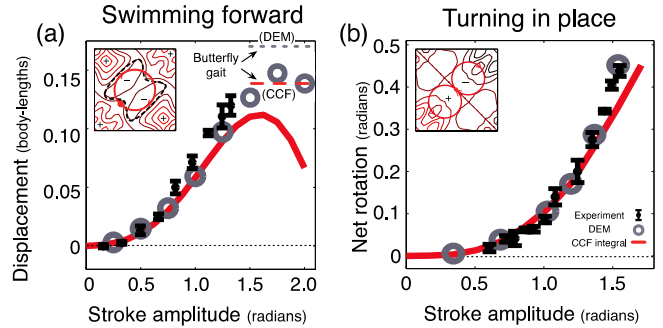}
\caption{Predictions of movement for the three-link swimmer: (A) CCF estimate of displacement compared to experimental and DEM results as a function of stroke amplitude (radius of the circle traced in the shape space). Dashed horizontal lines indicate displacements for the butterfly gait in DEM and RFT. The inset shows the circle gait overlaid on the $x^{b}$ CCF, along with a graphically-optimized ``butterfly" gaits (dashed path). (B) CCF estimate of net rotation for figure-eight gait of different stroke amplitudes, the radius of one circle of the figure-eight overlaid on the $\theta^{b}$ CCF in the inset. Frequency was 0.5 Hz in (A) and 0.17 Hz sin (B).}
\label{fig:predictions}
\end{figure}

By visualizing gaits as enclosures of area, the CCFs provide a comprehensive overview of how gait patterns interact with system constraints to produce net displacement. For example, we can explain the net forward displacement for the gait in Fig.~\ref{fig:vecheight}\subletter{a} by overlaying it on the $x^{b}$ CCF from Fig.~\ref{fig:geom}\subletter{A} (see inset of Fig.~\ref{fig:predictions}\subletter{A}). The large negative region at the center of the $x^{b}$ plot indicates that cycles in this region produce net $x$ translation relative to the starting frame, with clockwise (negatively oriented) cycles generating positive displacement. See SI Movie 1.

The CCFs also give insight into the relationship between net displacement and magnitude of the joint motions. For small amplitude circular gaits, where the sign is negative, the net displacement scales approximately quadratically with amplitude, tracking the rise in the enclosed area. At large amplitudes, the gait includes positive regions near the corners of the plot, reducing the area integral. These behaviors point to the optimal amplitude as that which encompasses as much of the central negative region as possible while avoiding the outlying positive regions. This geometric interpretation also suggests gaits, such as the butterfly in the inset of Fig.~\ref{fig:predictions}\subletter{A}, that produce more displacement than any circle by better conforming to the sign-definite regions; such a gait was created by fitting a polynomial curve to a set of points that were on the zero-set, symmetric about the origin, and avoided extreme joint angles~\footnote{We could not generate the large amplitude and butterfly gaits in experiment due to limitations on the robot's range of motion.}.

Agreement between theory, experiment and simulation is excellent at small amplitudes and good at larger amplitudes. However, while the CCFs qualitatively predict forward movement at the highest amplitudes, quantitative agreement is lacking. Previous work~\cite{Hatton:2010ASME,Hatton:2011IJRR} suggests a cause for this error. The velocity error is largest when joint angles are near the $\alpha_{1}=\alpha_{2}$ line and away from the origin, \emph{i.e.}, with links in a ``C" shape.  This error appears in Fig.~\ref{fig:vecheight}\subletter{C} as the system experiencing greater-than-expected velocity in the body $x$ and $y$ directions in the vicinity of this line (near $t=0.5$). In similar systems ({\em e.g.} the ``kinematic snake"~\cite{Shammas:2007,Hatton:2011IJRR}), this line is a kind of kinematic singularity, which forces constraints to slip. For granular-swimming, in which forces are inherently nonlinear, we postulate that the singularity makes the system reach different equilibria.


The CCFs facilitate study of movements that are relatively unexplored in swimming locomotion, for example, turning in place, see SI Movie 2. To design this gait, we observe that the $x$ component of the CCF is evenly symmetric around both the $\alpha_{1}=\alpha_{2}$ and $\alpha_{1}=-\alpha_{2}$ lines, the $y$ component is odd around $\alpha_{1}=\alpha_{2}$ and even around $\alpha_{1}=-\alpha_{2}$, and the $\theta$ component is even around $\alpha_{1}=\alpha_{2}$, but odd around $\alpha_{1}=-\alpha_{2}$. The figure-eight gait depicted in the inset of Fig.~\ref{fig:predictions}\subletter{B} will therefore produce a net rotation of the system---the two loops are in opposite directions in oppositely-signed regions, and so their effects add in $\theta$ while canceling in $x$ and $y$. In this gait, agreement between theory and experiment/DEM is excellent; we hypothesize that here the singularity described above is avoided.

The CCFs also facilitate comparison of swimming in different environments. The granular swimmer's CCFs in Fig.~\ref{fig:vecheight}\subletter{A} share a structure with those of other three-link swimming systems~\cite{Hatton:2010ASME,Hatton:2011IJRR}: a central well and bi-even symmetry in the $x^{b}$ function, and odd symmetry in the $y^{b}$ and $\theta^{b}$ functions.  Comparing them with equivalent plots for the low Re system~\cite{Hatton:2010ASME} (Fig.~\ref{fig:vecheight}\subletter{B}) highlights the differences between the environments. Most significantly, the $x^{b}$ function's magnitude is $1.5-2$ times larger for the granular swimmer than for the low Re system, in accord with the relatively larger $F_\perp$ in the granular medium. Interestingly, this relationship is reversed for the $\theta^{b}$ functions, indicating that turning gaits produce less rotation for swimmers in GM relative to those in low Re fluids. We note that the zero-sets we identify on the CCFs represent the optimal solutions discovered by the parametric optimization methods in~\cite{tamAhos}.


In summary, CCFs advance our understanding of locomotion by substituting geometric insight for laborious calculation. Since the technique requires only empirical force laws, we argue that this method lays the groundwork for geometric analysis of biological and robotic locomotion in environments that are not yet (and may never be) described by comprehensive equations of motion. We propose that the geometric insight gained by CCFs for locomotion will be analogous to insight into complex dynamical systems provided by low dimensional maps ~\cite{strogatzbookSHORT}. Future investigations will advance the linear model to represent dynamics near geometric singularities and incorporate effort-based distance metrics~\cite{Hatton:2011RSS} to describe locomotive efficiency. It will also be interesting to compare optimal locomotor strategies in different media combining the CCF analysis with effort metrics to identify maximally-efficient gaits. Finally, the use of continuous curvature modes~\cite{Hatton:2011RSS} will allow application of our framework to continuous systems like the biological sandfish, snakes that move within yielding substrates like loose soils, and even nematodes and spermatoza in complex biofluids~\cite{terAfau2010}.

{\em Acknowledgements--} We thank Andrew Masse and Ryan Maladen for experimental assistance, and Paul Umbanhowar and Lea Albaugh for discussion. Funding for RLH and HC provided by NSF Grant 1000389 and ARO Grant W911NF-11-1-0404; for YD, RDM, and DIG by The Burroughs Wellcome Fund, NSF PoLS Grant Nos. PHY-0749991 and PHY-1150760, ARO Grant W911NF-11-1-0514 and the ARL MAST CTA.
\bibliographystyle{apsrev4-1}
%

\end{document}